\documentclass[conference]{IEEEtran}
\usepackage{amsmath,amssymb,amsfonts}
\usepackage{booktabs}
\usepackage[normalem]{ulem}
\usepackage{array}
\usepackage{graphicx}
\usepackage{subfig}
\usepackage{setspace}
\usepackage{cite}
\usepackage{multicol}
\usepackage{float}
\usepackage{multirow}

\newcommand{\MYfooter}{\smash{
		\hfil\parbox[t][\height][t]{\textwidth}{}\hfil\hbox{}}}
\makeatletter
\def\ps@IEEEtitlepagestyle{%
	\def\@oddfoot{\MYfooter{}}%
	\def\@evenfoot{\MYfooter}}	
\makeatother
\usepackage{array}
\newcolumntype{L}[1]{>{\raggedright\let\newline\\\arraybackslash\hspace{0pt}}m{#1}}
\newcolumntype{C}[1]{>{\centering\let\newline\\\arraybackslash\hspace{0pt}}m{#1}}
\newcolumntype{R}[1]{>{\raggedleft\let\newline\\\arraybackslash\hspace{0pt}}m{#1}}

\begin{document}
\title{BRIGHTNESS: Leaking Sensitive Data from Air-Gapped Workstations via Screen Brightness}
\author{\IEEEauthorblockN{Mordechai Guri\IEEEauthorrefmark{1}, Dima Bykhovsky\IEEEauthorrefmark{1}\IEEEauthorrefmark{3}, Yuval Elovici}
	\IEEEauthorblockA{\IEEEauthorrefmark{1}Department of Software and Information Systems Engineering, Ben-Gurion University of the Negev, Israel\\
		Email: gurim@post.bgu.ac.il}
	\IEEEauthorblockA{\IEEEauthorrefmark{3}Department of Electrical and Electronics Engineering, Shamoon College of Engineering, Israel \\ demo video: https://cyber.bgu.ac.il/advanced-cyber/airgap \\}
	} 
\maketitle

\begin{abstract}
Air-gapped computers are systems that are kept isolated from the Internet since they store or process sensitive information. 
In this paper, we introduce an optical covert channel in which an attacker can leak (or, exfiltlrate) sensitive information from air-gapped computers through manipulations on the screen brightness. This covert channel is invisible and it works even \textit{while} the user is working on the computer. Malware on a compromised computer can obtain sensitive data (e.g., files, images, encryption keys and passwords), and modulate it within the screen brightness, invisible to users. The small changes in the  brightness are invisible to humans but can be recovered from video streams taken by cameras such as a local security camera, smartphone camera or a webcam. We present related work and discuss the technical and scientific background of this covert channel. We examined the channel's boundaries under various parameters, with different types of computer and TV screens, and at several distances. We also tested different types of camera receivers to demonstrate the covert channel. Lastly, we present relevant countermeasures to this type of attack. 
\end{abstract}

\section{Introduction}
Despite the existence of security measures such as intrusion detection systems (IDS), firewalls and AV programs - attackers are finding new vulnerabilities and ways to infiltrate target networks. Even networks that are completely disconnected from the Internet can be compromised by motivated adversaries using complex attack vectors. While breaching such systems has been shown to be feasible in recent years, exfiltration of data from systems without networking or physical access is still considered a challenging task. Electromagnetic, acoustic and thermal covert exfiltration channels have been examined in the last twenty years. Optical exfiltration techniques have been studied as well \cite{Loughry2002}. However, most of these methods are visible (e.g., not covert) and assume the \textit{absence} of people in the environment.

In this paper, we propose an optical covert channel which relies on the limitations of human vision. Technically speaking, visible light represents a limited range of electromagnetic radiation, which is sensed and perceived by the human visual system. Intentional leakage of sensitive data through the visible light via a standard LCD screen is futile, since by definition it may be detected by humans who see the display. Our covert channel exploits the limits of human visual brightness perception in order to conceal sensitive data, invisible to the naked eye, on the LCD screen. 

\section{Related Work}
Leaking data from air-gapped systems via covert communication methods has been explored in the last twenty years. The covert channels studied are electromagnetic, magnetic, electric, acoustic, thermal, and optical. Back in 1998 \cite{Kuhn1998} researchers discuss the concept of software based TEMPEST attack, which employs electromagnetic emanation from LCD screen. AirHopper \cite{Guri2017c} is an attack aimed at exfiltrating data from isolated networks via radio frequencies in the FM broadcasting bands (87.5 - 108.0 MHz). The signal are received by FM radio chip in a standard smartphone. Electromagnetic covert channels are discussed in \cite{guri2016usbee,Guri2015,Kasmi2016,Yang2017,Zhou2019b ,guri2018lcd} and newer magnetic convert channels discussed in  \cite{guri2019odini,guri2018magneto}. 
Hanspach and Goetz \cite{Hanspach2013} present a method for near-ultrasonic covert networking using speakers and microphones. Fansmitter and Diskfiltration \cite{guri2016fansmitter,guri2017acoustic} are another methods of acoustic data exfiltration from computers without loudspeakers. BitWhisper \cite{Guri2014} demonstrates a covert communication channel between adjacent air-gapped computers by using their heat emissions and built-in thermal sensors. In 2018, Guri et al presented PowerHammer \cite{2018powerhammer}, a method to exfiltrate data from air-gapped computers through power lines. Other types of air-gap covert channels based on acoustic \cite{deshotels2014inaudible,hanspach2014covert,carrara2014acoustic,Krishnamurthy2018}, optical \cite{guri2016optical,loughry2002information,lopes2017platform,Guri2017,Loughry2018,Bak2019,Zhou2019,Guri2019} and thermal \cite{guri2015bitwhisper} emissions have also been investigated.

In 2002 researcher \cite{Loughry2002} discuss the threat of data leakage through optical emanations from LEDs. They manipulates the keyboard LEDs to modulate data which was received by a camera. Using the keyboard LEDs for covert channel discussed was explored in 2019 with modern keyboard and smartphones \cite{guri2019ctrl}. Recently researchers discuss the threat of modulating information via the routers and LAN switches LEDs \cite{guri2018xled}. Brasspup \cite{Sar13} demonstrated how to hide images in a modified LCD screen but his method requires hardware modification. The term 'shoulder surfing' refers to a malicious insider or visitor looking at the screen or carrying a camera. Another threat is an exploited surveillance camera. The visitors or cameras are obtaining private data such as credit card (CC) numbers, passwords and PIN codes. With our method, the presence or absence of the user is not required, since the attacker may leak the sensitive information at any time in a stealth way. 

\section{Attack Model}
\begin{figure}[b]
	\centering
	\includegraphics[width=3.32in,height=1.99in]{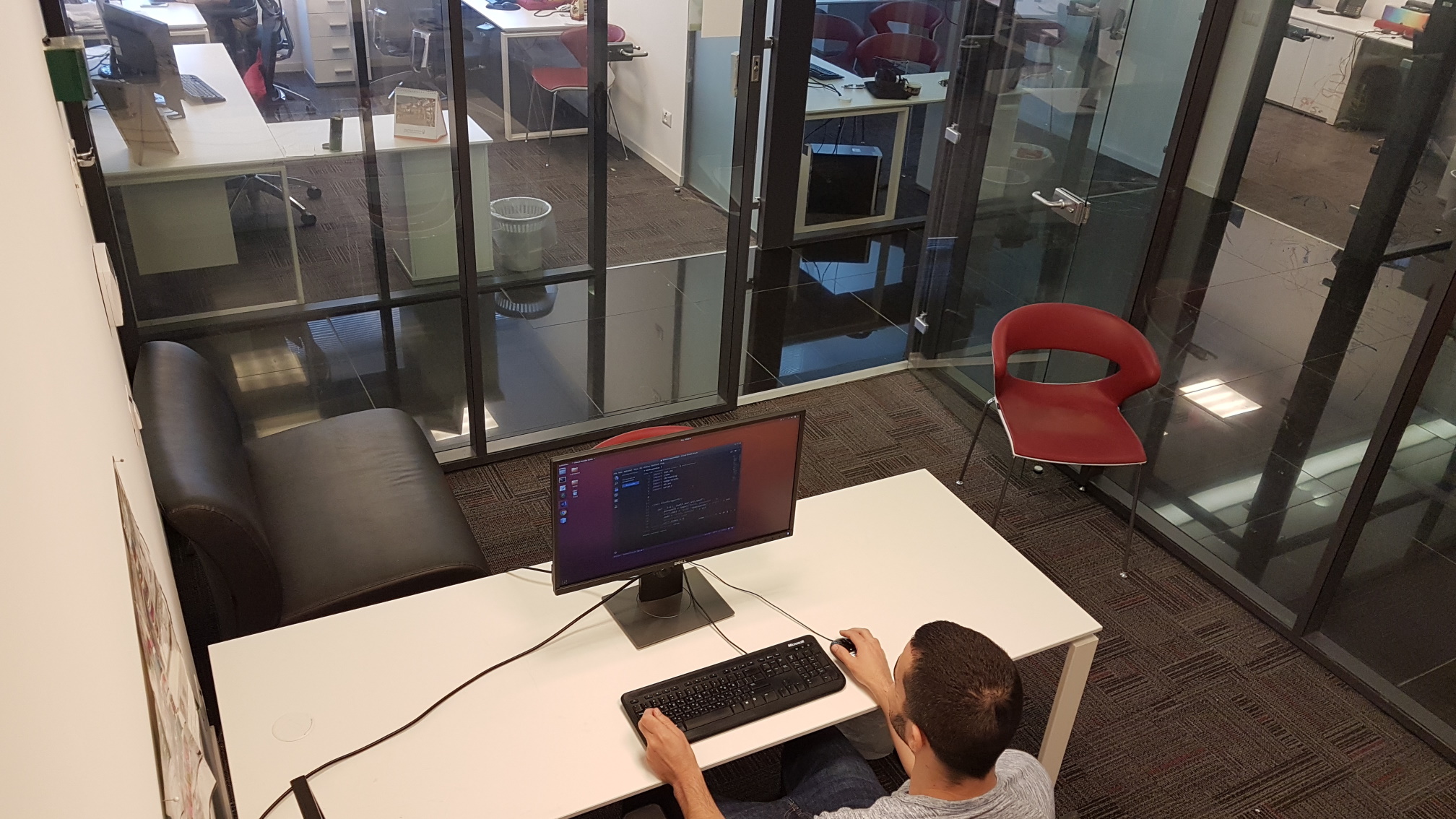}
	\caption{Sensitive data is exfiltrated from the computer, modulated within the screen brightness, invisible and unbeknownst to the user. The video stream is recorded by a local security camera.}
	\label{fig:1}
\end{figure}

	At the first stage of the attack, the target network is infected with a malware. Infiltration of air-gapped network can be done in a case of motivated and capable adversaries \cite{Osnos2017,Beatingt16:online}. At the second stage, the malware collects sensitive information from the computer (e.g., documents). It then encodes it as a stream of bytes and modulates it on the screen, using small changes in the screen brightness that is invisible to humans. The third stage of the attack involves a camera which takes video recordings of the compromised computer's display. Attackers then access the recorded video stream and reconstruct the sensitive information by using image processing techniques. There are two attack scenarios which are relevant for this optical covert channel. The 'malicious insider' attack \cite{Tri15} (also known as 'evil maid' attack \cite{Tec}) in which a person with a camera can be within the compromised computer's line of sight but does not necessarily have network access. A  compromised local camera (e.g., surveillance camera) that the attacker has access to. A sample attack scenario is depicted in Fig.~\ref{fig:1}. In this scenario sensitive data (e.g., encryption key) is covertly modulated onto the computer screen brightness. It then projected on the screen either when the user is absent or while the user is working on the computer. The modulated data is then reconstructed from the video stream of a local security camera using video processing techniques.




Notably, threat models in which the attacker must be in close proximity of the emanating device are common in a variety of covert channels \cite{Loughry2002,Las09,Guri2015,Morkel2005}.

\section{Technical Background}
\begin{figure}[h]
	\centering
	\includegraphics[width=\linewidth]{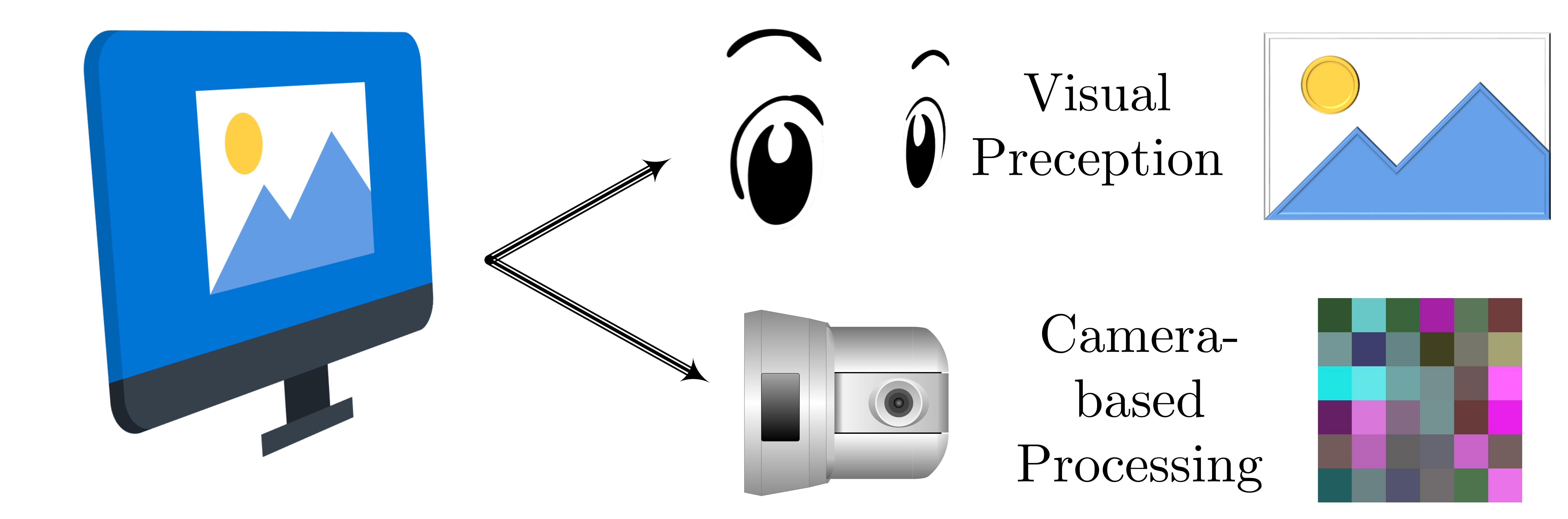}
	\caption{Illustration of covert D2C principle: the same frame on the screen has both high-quality visually perceptible image data and imperceptible covert message.}
	\label{fig:illustration}
\end{figure}

Display-to-camera (D2C) communication is a subject of significant recent interest. In this communication, a camera is used for both accessing scene elements and capturing imperceptible machine-interpretable data. The main application of D2C is to provide legitimate covert channels for multimedia services \cite{Yuan2012,Wang2014,Li2015,Kim2015,Jung2018,Wang2017}. An illustration of D2C is presented in Fig.~\ref{fig:illustration}. 

Numerous techniques were proposed for D2C. 
In \cite{Yuan2012}, it was proposed to apply the Laplacian Pyramid method based watermark technique that embeds data in an image frame. 
Wang et al \cite{Wang2014} used complementary image frames with data blocks added to the original images, while preserving them to be imperceptible to the human eye.
Li et al \cite{Li2015} used alpha channel manipulations. 
In \cite{Kim2015,Jung2018}, a frequency domain communication method inspired by orthogonal frequency-division multiplexing (OFDM) modulation was applied.
In \cite{Wang2017}, pixels or blocks of pixels are modulated by a spatial visual modulation scheme.

\subsection{Challenges}
\begin{figure}[h]
	\centering
	\includegraphics[width=.45\linewidth,page=3]{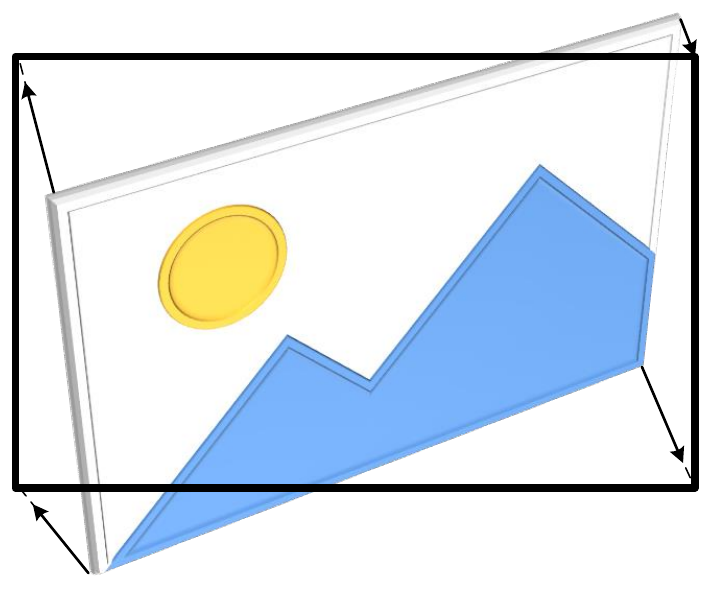}
	\caption{Illustration of inverse affine transformation that is required for correct processing of communication data.}
	\label{fig:challenges}
\end{figure}

The implementation of covert D2C communication requires minimum changes of the displayed information such that it appears unchanged to the human eye. On the other hand, the communication scheme must be immune to the effects of camera geometry, such as the scale, angle rotation and optical distortion. For example, the required inverse affine transformation of the captured image is illustrated in Fig. \ref{fig:challenges}. 
While, in general, all mentioned effects may be effectively mitigated by sufficient image processing, they have an inevitable influence on communication performance and the required computation complexity, thus making it challenging to provide a robust communication scheme with an acceptable bit-rate and bit-error rate.

\subsection{Performance}\label{sec:performance}
The main parameter that influences communication performance is the received optical power and the resulting SNR. The power, in turn, depends on: distance to the display, misalignment between the camera and the display, affine transformation of the displayed image (also termed perspective distortion), optical zoom of the camera, display contrast ratio and brightness. 

\begin{figure}[h]
	\centering
	\includegraphics[width=.75\linewidth,page=2]{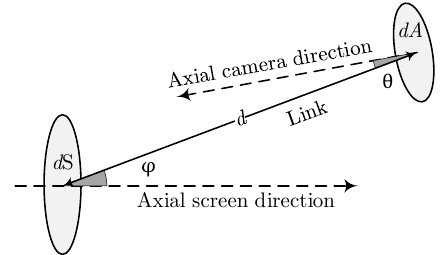}
	\caption{D2C link geometry that includes axial misalignment of both a display and a camera.}
	\label{fig:performance}
\end{figure}

Optical channel gain analysis is based on the geometric parameters outlined in Fig. \ref{fig:performance}. The distance between the display and the camera is $d$, the axial misalignment of the display is $\phi$ and the axial misalignment of the camera is $\theta$. The relation between power emitted from display element $dS$ towards camera element $dA$ is given by \cite{Saleh2007}
\begin{equation}\label{eq:optical_link_budget}
dh = \int\limits_{screen}\int\limits_{\substack{camera\\objective}}\frac{1}{\pi d^2}\cos(\phi)\cos(\theta)dAdS.
\end{equation}
This rigorous expression requires integration over the display area and the camera objective area of the channel gain above.
This expression stresses the dependence of the gain on distance, $\sim d^{-2}$, and axial misalignment, $\sim\cos(\cdot)$.

The main image processing aspect is related to the affine transformation of the image.
The reverse affine transformation (Fig. \ref{fig:challenges}) is not only computationally intensive, but also produces inevitable image quality degradation, thus degrading communication performance \cite{Kim2015}. 

The optical zoom of the camera physically defines the number of detector pixels that are used for the imaging of display information. Obviously, higher zoom results in better communication performance and enables a higher bit-rate and/or BER. Note that the display contrast ratio and brightness are directly related to the transmitted optical power of the communication signal. Higher brightness thresholds yield higher quality of the communication channel.

\section{Data Communication}\label{sec:brightness}
\begin{figure}
	\centering
	\includegraphics[width=.65\linewidth,page=2]{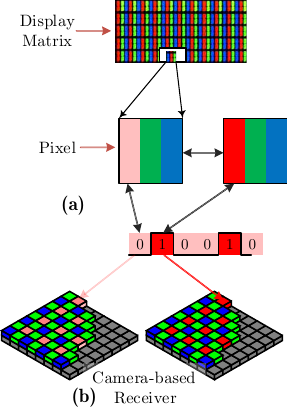}
	\caption{(a) The signal is modulated by imperceptible changes of one of the RGB components. In this figure, small R(ed) color changes are used for modulation. (b) Camera-based receiver is used for signal detection.}
	\label{fig:red_screen}
\end{figure}

In LCD screens each  pixel presents a combination of RGB colors which produce the required compound color. An illustration of the RGB principle is presented in Fig. \ref{fig:red_screen}(a).

In the proposed modulation, the RGB color component of each pixel is slightly changed. These changes  are invisible, since they are relatively small and occur fast, up to the screen refresh rate. Moreover, the overall color change of the image in the screen is invisible to the user. The modulation process is outlined in Fig. \ref{fig:red_screen}(b).

In the general case we use $M$-level amplitude-shift keying (M-ASK). In this modulation, different lighting levels are used to represent a \textit{symbol} that includes $\left\lfloor{\log_2\left(M\right)}\right \rfloor$ bits. Typically, the value of $M$ is chosen to be a power of two. Each symbol has the same duration, $T$, such that the resulting bit-rate is given by
$R=\left\lfloor{\log_2\left(n\right)}\right \rfloor T$ bit/sec. The special case of $M=2$ is termed on-off keying (OOK). 

\begin{figure}
	\centering
	\includegraphics[width=0.95\columnwidth]{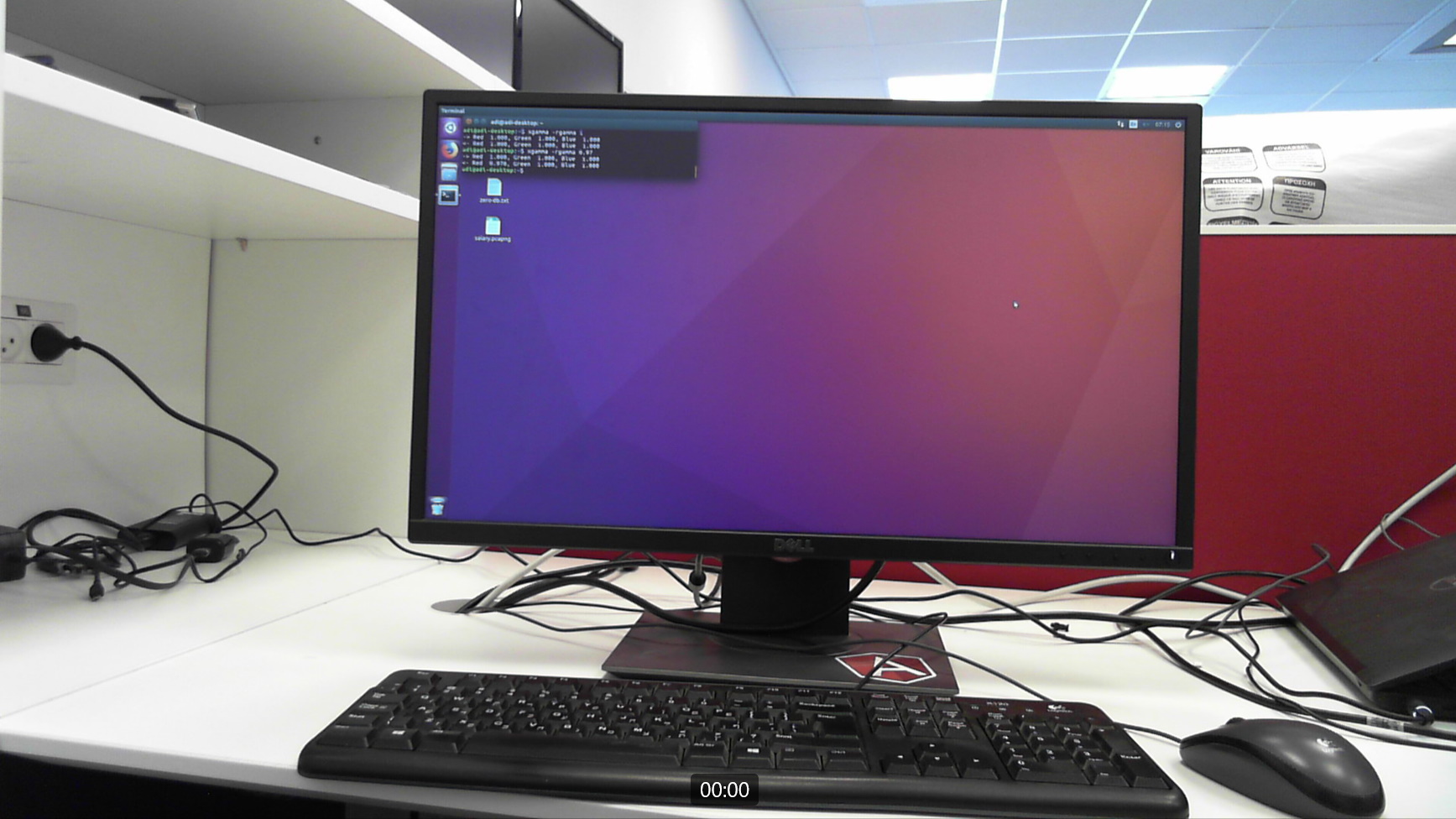}
	\includegraphics[width=0.95\columnwidth]{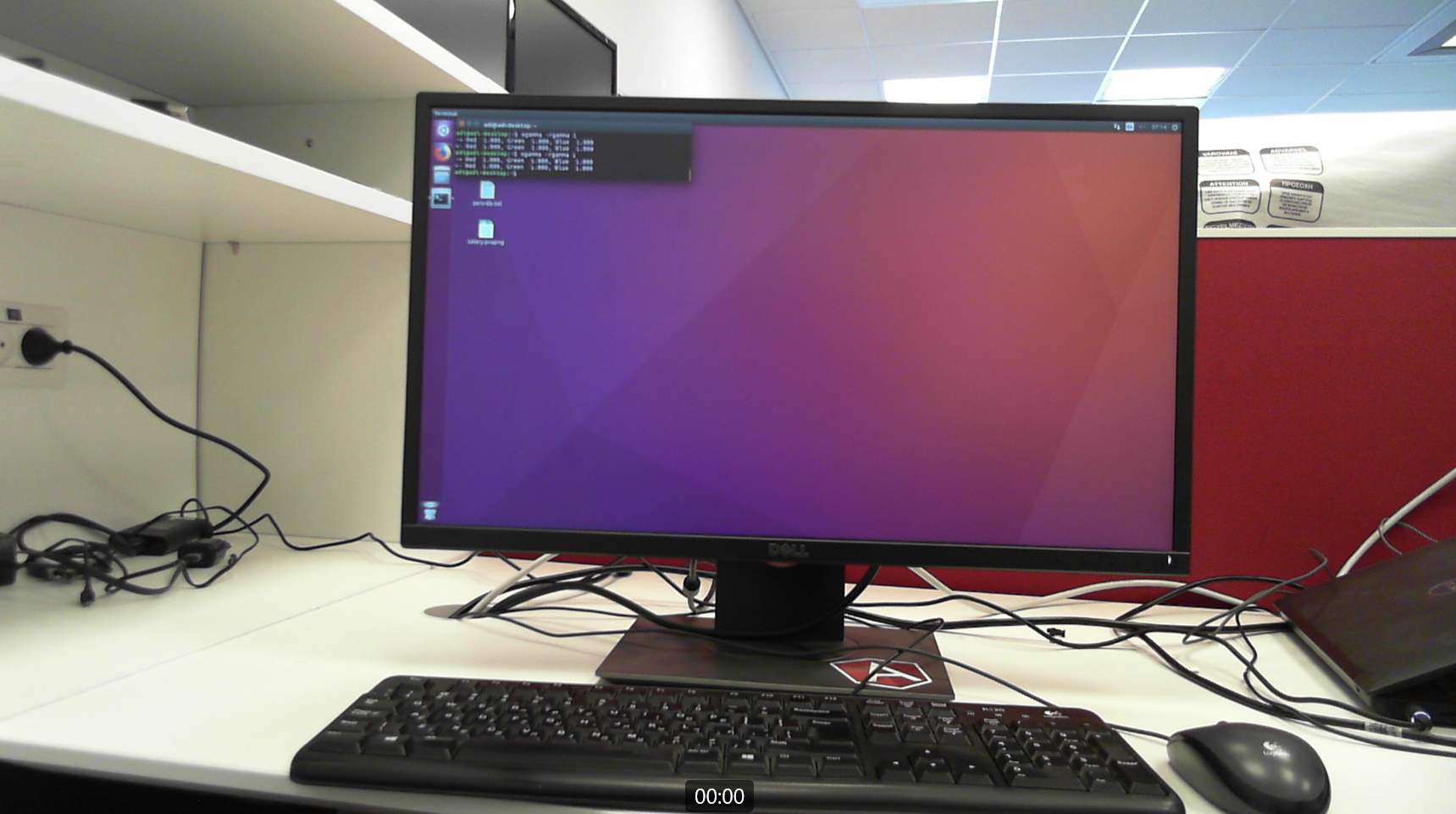}
	\caption{The ASK modulation in a video stream. The `1' and `0' are modulated in the brightness of the top and bottom screens, respectively.}
	\label{fig:red2}
\end{figure}

\begin{figure}
	\centering
	\includegraphics[width=.85\linewidth]{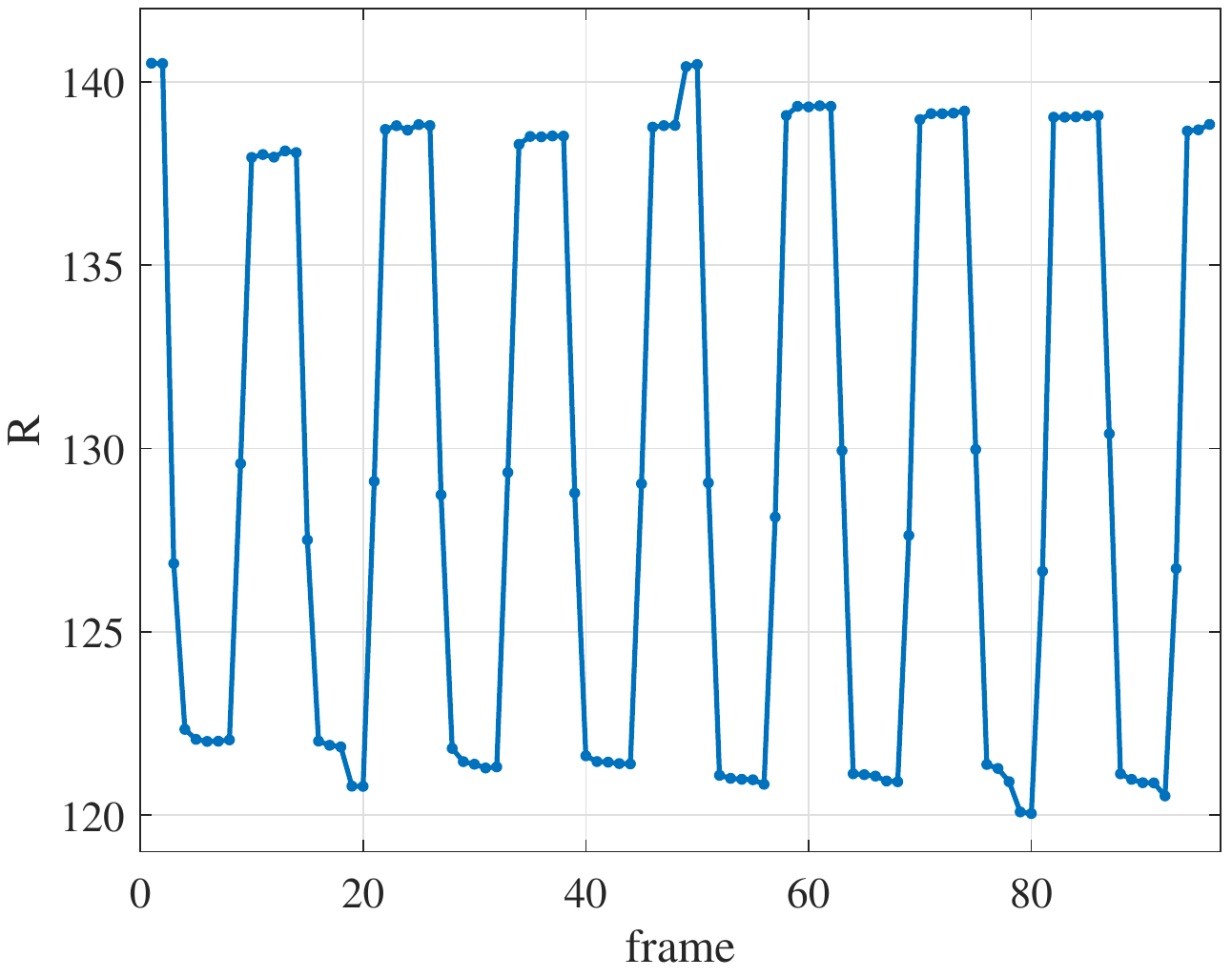}
	\caption{Signal `1010101010101010' modulated by color change as acquired by a security camera.}
	\label{fig:red_signal}
\end{figure}

An example of the signal modulated by 3\% changes in the red color component is presented in Fig. \ref{fig:red2}. It shows the ASK modulation in two frames from the video stream. The `1' and `0' values are modulated in the brightness of the top and bottom screens, respectively. The analysis of the stream is presented in Fig.~\ref{fig:red_signal}. In this case a bit sequence of `1010101010101010' was exfiltrated from a 19" screen at a bitrate of 5 bit/sec. It was captured by a local camera located at a distance of 6 meter from the screen.

\begin{figure}
	\centering
	\includegraphics[width=\linewidth, page=2]{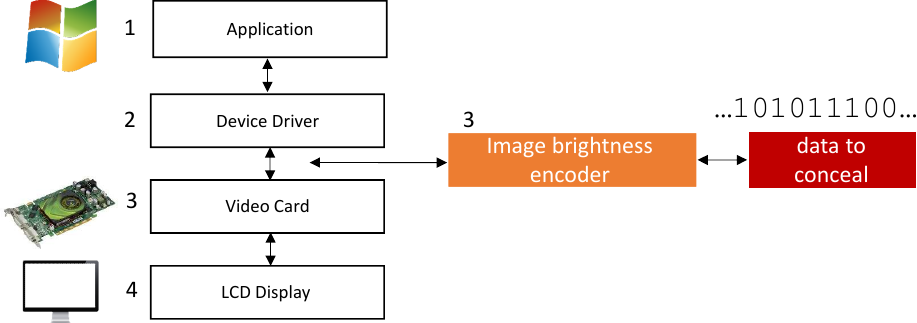}
	\caption{The malware architecture}
	\label{fig:arch}
\end{figure}

The malware architecture is presented in Fig. \ref{fig:arch}. The image brightness encoder is a device driver which intercepts the screen buffer. It modulates the data in ASK by modifying the brightness of the bitmap according to the current bit (`1' or `0'). It changes the RGB component of every pixel by a given amount and forwards it to the video card.

\section{Evaluation}
We evaluate the covert communication channel as a function of different distances and bitrates.

\subsection{Experimental setup}
For the transmission we used two display screens; (1) Dell 24" PC Monitor (P2417H), and (2) Samsung 40" LED TV (UA40B6000). For the reception we tested three types of cameras: (1) a professional security camera, (2) a webcam, and (3) a smartphone camera. The camera models and the other details are summarized in Table \ref{tab:cameras}. For decoding the videos we used OpenCV, which is open-source computer vision library that focuses on real-time video processing for academic and commercial use. We developed a C program that receives the video as an input and calculates the frame brightness and illumination amplitudes to an output file for further MATLAB processing.

%
\begin{table*}[]
	\centering
	\caption{Evaluation of brightness based covert channel with different receivers}
	\label{tab:cameras}
	\begin{tabular}{@{}lllcc@{}}
		\toprule
		\# & Name            & Model                                 & Distance (m) & Bit-rate (bps) \\ \midrule
		1  & Security camera & Sony SNC-DH120 IPELA Minidome 720P HD & 1-9          & 5-10           \\
		2  & Webcam          & Microsoft Lifecam Studio              & 1-9          & 5-10           \\
		3  & Smartphone      & Samsung Galaxy S7                     & 0.3-1.5      & 1              \\ \bottomrule
	\end{tabular}
\end{table*}

Our experiments show that the best communication performance was for adapting the red color component. In this modulation, we changed the red color of each pixel by a maximum threshold of 3\%. The changes are invisible to humans but can be reconstructed from a recorded video stream. For the security camera and the webcam we could reach bit-rates of 10 bps and a maximal communication distance of 9 meters. The resulting bit-error rate (BER) was 0\% for all experiments.  

Note, the communication distances in both communication schemes were limited by the available indoor environment. Large distances are possible, as discussed in the following section.

\subsection{Distance Analysis}
\begin{figure}[b]
	\centering
	\includegraphics[width=.85\linewidth]{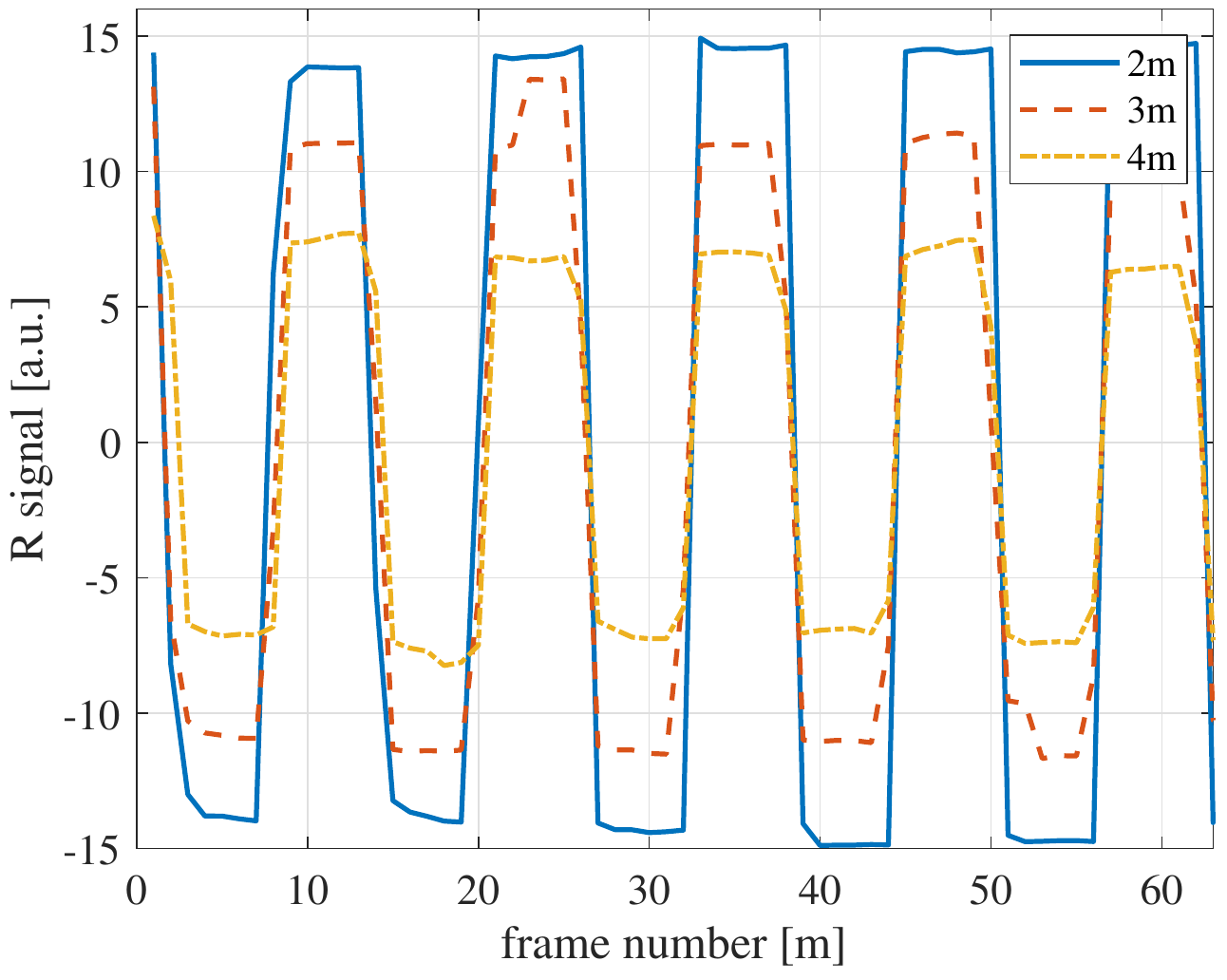}
	\caption{Received signal examples from different distances.}
	\label{fig:distance_analysis_visible}
\end{figure}

An example of the signal decrease for a red color modulated signal as a function of distance is presented in Fig. \ref{fig:distance_analysis_visible}. It shows the red component as extracted from the video stream of modulated `1' and `0' values at three distances.

Theoretically, the received signal is inversely proportional to the squared distance \cite{Saleh2007}:
\begin{equation}\label{eq:distance}
\mathrm{signal}\sim \frac{1}{d^2}.
\end{equation}

\begin{figure}
	\centering
	\includegraphics[width=.75\linewidth]{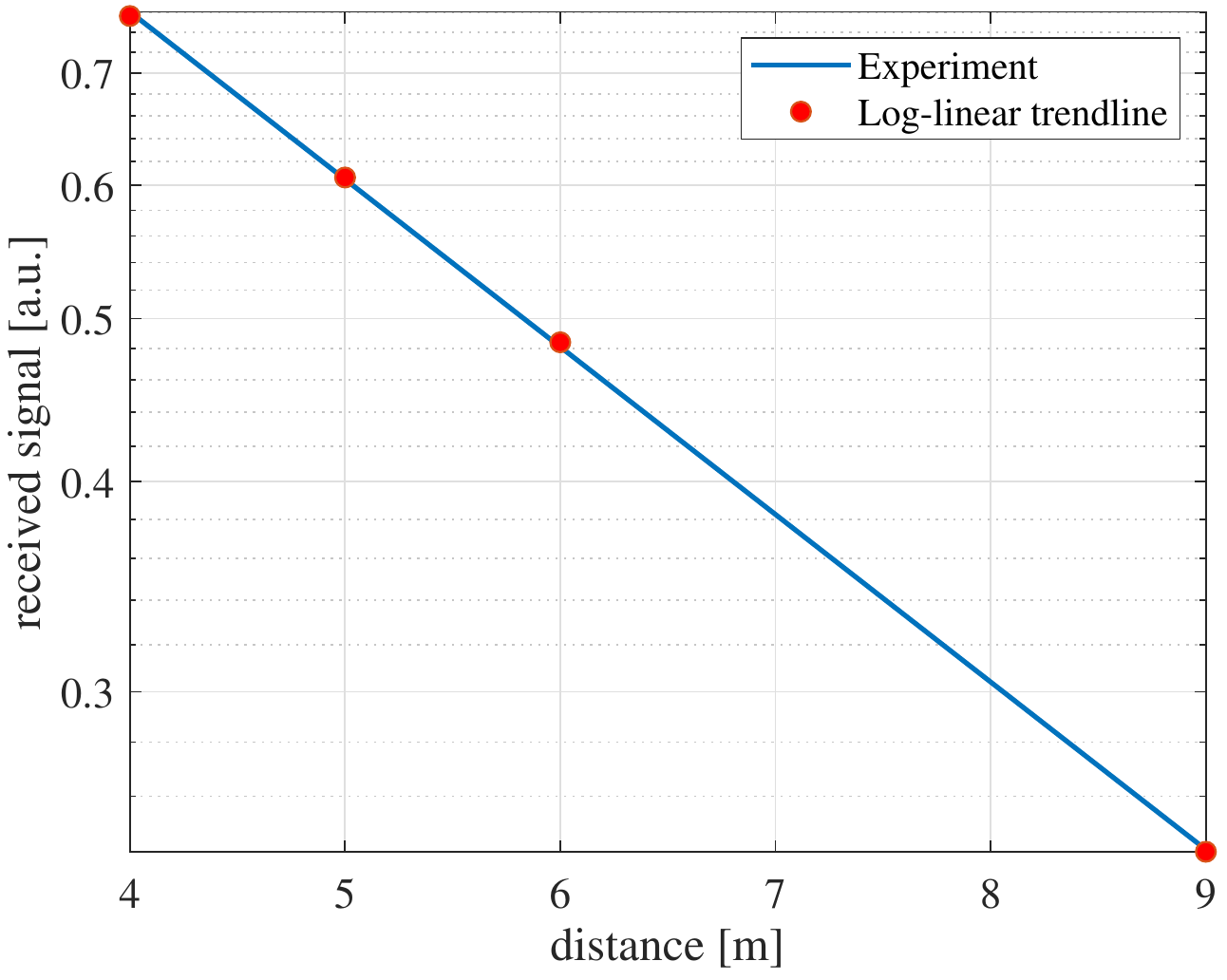}
	\caption{Received signal is inversely proportional to the squared distance.}
	\label{fig:distance_analysis}
\end{figure}

In order to verify this dependence, the experiment of brightness-modulated communication was repeated for different distances and the results were analyzed. The analysis of signal variability is presented in Fig. \ref{fig:distance_analysis} and clearly shows that the theoretical dependency (Eq. \eqref{eq:distance}) holds for the experimental results. Note, the \textit{y}-axis of the figure has a logarithmic scale, so the resulting linear trend-line demonstrates squared distance dependency.

\section{BER Analysis}\label{sec:ber_analysis}
The bit error rate (BER) of the communication channel be evaluated as 
\begin{equation}
p_e = p(0)p(1|0) + p(1)p(0|1),
\end{equation}
where $p(0)$ and $p(1)$ are the probabilities of transmitted `0' or `1' respectively and $p(1|0)$ and $p(0|1)$ are correspondent conditional error probabilities \cite{Proakis5th}.
Using central limit theorem approximation, the received signals may be modeled by $s_0 \sim N(\mu_0,\sigma_0^2)$ and $s_0 \sim N(\mu_1,\sigma_1^2)$ with the corresponding conditional probabilities 
\begin{subequations}
	\begin{align}
	p(1|0) &= Q\left(\frac{thr-\mu_0}{\sigma_0}\right) \\
	p(0|1) &= Q\left(\frac{\mu_1-thr}{\sigma_1}\right),
	\end{align}
\end{subequations}
where $\gtrless\mathrm{thr}$ is the decision threshold for the received signal and 
\(Q(x) = \frac{1}{\sqrt{2\pi}}\int_{x}^{\infty}\exp\left(-\frac{x^2}{2}\right).\)
For simplifying conditions of $p(0) = p(1) = 1/2, \sigma_0=\sigma_1=\sigma$ and applying the optimal $thr=(\mu_1-\mu_0)/2$ value, the resulting BER expression is given by
\begin{equation}
p_e = Q\left(\frac{\mu_1-\mu_0}{2\sigma}\right).
\end{equation}
While $\sigma$ value may be considered as constant, the difference $\mu_1-\mu_0$ reduces with distance as outlined in the evaluation.

\section{Scientific discussion}
In this section we present the scientific background regarding human visual perception limitations which facilitate the success our optical covert channel. The ability of humans to resolve blinking images and brightness perception are discussed in \cite{Whe1,Ada1,Yam12,Placeholder1}. Physiological aspects of human color vision are discussed further by Gouras \cite{Gou91}. Coren et al also provide a general discussion of the human visual system \cite{Cor99}, along with details regarding the perception of brightness and darkness \cite{Cor99,Cor991}, lightness constancy \cite{Cor99}\cite{Cor992}, and temporal properties of the visual system \cite{Cor99,Cor993}.

\subsection{Brightness and darkness perception}
The level of ambient (environmental) light is known to affect visual perception, including the perception of brightness \cite{Cor99,Cor991}. In fact, the human visual system consists of photopic or daylight vision, which includes the perception of color, and scotopic or twilight vision. In the human retina, two separate types of cells (cones and rods) are responsible for daylight and twilight vision: cones are associated with photopic vision, while rods are associated with scotopic vision. The sensitivity of the visual system gradually adapts as one move to a darker or brighter environment. Consequently, our experiments are performed under a controlled level of ambient light. Also, subjects are given some time to adapt to the laboratory’s level of ambient light. It is also believed that human perception of relative brightness and darkness involves two separate systems \cite{Cor99,Cor991}. 

Concerning the duration of the blinking image, particularly with low levels of illumination, increasing the duration can increase the likelihood that the stimulus will be detected, a phenomenon known as Bloch's law \cite{Cor99,Cor991}. Concerning perception of flickering light, the retinal receptors in the human eye can resolve up to several hundred cycles per second (cps). However, the sensitivity of neurons in the primary visual cortex to flickers is much lower \cite{Cor99,Cor993}. The critical fusion frequency (CFF) is used to measure subjects' discrimination between steady and flickering light. This measure varies between 10 cps and 60 cps (exposure time between 50 ms and 8.3 ms, respectively). The CFF varies based on several factors, including the current level of light/dark adaptation, the intensity of the light, the distance from the fovea, and the wavelength composition of the light. Consequently, our experiments are performed under controlled values for those factors. In the human retina, separate ganglion cells are responsible for sustained (steady) light and transient (flickering) light (see also \cite{Cor99}). Interestingly, it has been demonstrated \cite{Cor99,Cor993} that low contrast flashes and equiluminant chromatic (color) flashes activate different pathways. In this research, we are particularly interested in low contrast flashes of gray tones.

\section{Countermeasures}
The countermeasures for this optical attack can be categorized to prevention and detection. Preventive countermeasures include organizational policies aimed to restrict the accessibility of sensitive computers by placing them in secured areas ('zones') where only authorized staff may access them. In addition any sort of cameras (including smartphone and wearable cameras) may be prohibited  within the perimeter of certain restricted areas. Note that the surveillance camera itself may be infected with a malware. Another technological countermeasure consists of a polarized film which covers the screen. The user gets a clear view while humans and cameras at a distance would view a darkened display. Detection countermeasures may include monitoring of the sensitive computer for the presence of suspicious display anomalies at runtime. However, detection mechanisms within the operating systems are considered untrusted since they can be evaded by  malware (e.g., rootkits) at the user and kernel levels.
A trusted monitoring can be achieved by taking videos of the computer’s display and searching for hidden brightness change patterns. The detection can be done by a camera-based receiver. Since camera sensors are based on RGB optical filter arrays, the signal can be detected by one of the sensor components. However, practical implementation of camera based monitoring seems nontrivial in the wide scale due to the resources and maintenance it requires.


\section{Conclusion}
In this paper we present an optical covert channel in which data is concealed on the LCD screen brightness, but is invisible to users. We presented a malware scheme that can exfiltrate sensitive data from isolated (air-gapped) computers. The attack model consists of (a) contaminating the target network, a task which has been demonstrated to be within the capabilities of a modern advanced persistent threat (APT), and (b) using a camera to take videos of the computer’s display, a task that can be performed by a malicious insider or visitor, or by exploiting a surveillance camera. We exploit the limitations of bare human vision, concerning brightness perception, using sufficiently low values of contrast between the brightness levels. Consequently, the current results demonstrate the feasibility of our covert channel, while outlining its boundaries. Notably, this kind of covert channel is not monitored by existing data leakage prevention systems.

\bibliographystyle{ieeetran}
\bibliography{cyber_dima4}

\begin{thebibliography}{10}
\providecommand{\url}[1]{#1}
\csname url@samestyle\endcsname
\providecommand{\newblock}{\relax}
\providecommand{\bibinfo}[2]{#2}
\providecommand{\BIBentrySTDinterwordspacing}{\spaceskip=0pt\relax}
\providecommand{\BIBentryALTinterwordstretchfactor}{4}
\providecommand{\BIBentryALTinterwordspacing}{\spaceskip=\fontdimen2\font plus
\BIBentryALTinterwordstretchfactor\fontdimen3\font minus
  \fontdimen4\font\relax}
\providecommand{\BIBforeignlanguage}[2]{{%
\expandafter\ifx\csname l@#1\endcsname\relax
\typeout{** WARNING: IEEEtran.bst: No hyphenation pattern has been}%
\typeout{** loaded for the language `#1'. Using the pattern for}%
\typeout{** the default language instead.}%
\else
\language=\csname l@#1\endcsname
\fi
#2}}
\providecommand{\BIBdecl}{\relax}
\BIBdecl

\bibitem{Loughry2002}
J.~Loughry and D.~A. Umphress, ``Information leakage from optical emanations,''
  \emph{ACM Transactions on Information and System Security (TISSEC)}, vol.~5,
  no.~3, pp. 262--289, 2002.

\bibitem{Kuhn1998}
M.~G. Kuhn and R.~J. Anderson, ``Soft tempest: Hidden data transmission using
  electromagnetic emanations.'' in \emph{Information hiding}, vol. 1525.\hskip
  1em plus 0.5em minus 0.4em\relax Springer, 1998, pp. 124--142.

\bibitem{Guri2017c}
M.~Guri, M.~Monitz, and Y.~Elovici, ``Bridging the air gap between isolated
  networks and mobile phones in a practical cyber-attack,'' \emph{ACM
  Transactions on Intelligent Systems and Technology (TIST)}, vol.~8, no.~4,
  p.~50, 2017.

\bibitem{guri2016usbee}
------, ``{USBee}: Air-gap covert-channel via electromagnetic emission from
  {USB},'' in \emph{14th Annual Conference on Privacy, Security and Trust
  (PST)}.\hskip 1em plus 0.5em minus 0.4em\relax IEEE, 2016, pp. 264--268.

\bibitem{Guri2015}
M.~Guri, A.~Kachlon, O.~Hasson, G.~Kedma, Y.~Mirsky, and Y.~Elovici, ``Gsmem:
  Data exfiltration from air-gapped computers over {GSM} frequencies.'' in
  \emph{USENIX Security Symposium}, 2015, pp. 849--864.

\bibitem{Kasmi2016}
C.~Kasmi, J.~Lopes~Esteves, and P.~Valembois, ``Air-gap limitations and bypass
  techniques: “command and control” using smart electromagnetic
  interferences,'' \emph{The Journal on Cybercrime \& Digital Investigations},
  vol.~1, no.~1, Dec. 2016, {Botconf} 2015, Paris.

\bibitem{Yang2017}
Z.~Yang, Q.~Huang, and Q.~Zhang, ``Nicscatter: Backscatter as a covert channel
  in mobile devices,'' in \emph{Proceedings of the 23rd Annual International
  Conference on Mobile Computing and Networking}, ser. MobiCom '17.\hskip 1em
  plus 0.5em minus 0.4em\relax New York, NY, USA: ACM, 2017, pp. 356--367.

\bibitem{Zhou2019b}
Z.~{Zhou}, W.~{Zhang}, and N.~{Yu}, ``{Data Exfiltration via Multipurpose RFID
  Cards and Countermeasures},'' \emph{arXiv e-prints}, p. arXiv:1902.00676,
  Feb. 2019.

\bibitem{guri2018lcd}
M.~Guri and M.~Monitz, ``Lcd tempest air-gap attack reloaded,'' in \emph{2018
  IEEE International Conference on the Science of Electrical Engineering in
  Israel (ICSEE)}.\hskip 1em plus 0.5em minus 0.4em\relax IEEE, 2018, pp. 1--5.

\bibitem{guri2019odini}
M.~Guri, B.~Zadov, and Y.~Elovici, ``Odini: Escaping sensitive data from
  faraday-caged, air-gapped computers via magnetic fields,'' \emph{IEEE
  Transactions on Information Forensics and Security}, 2019.

\bibitem{guri2018magneto}
M.~Guri, A.~Daidakulov, and Y.~Elovici, ``Magneto: Covert channel between
  air-gapped systems and nearby smartphones via cpu-generated magnetic
  fields,'' \emph{arXiv preprint arXiv:1802.02317}, 2018.

\bibitem{Hanspach2013}
M.~Hanspach and M.~Goetz, ``On covert acoustical mesh networks in air,''
  \emph{Journal of Communications}, vol.~8, no.~11, pp. 758--767, Nov. 2013.

\bibitem{guri2016fansmitter}
M.~Guri, Y.~Solewicz, A.~Daidakulov, and Y.~Elovici, ``Fansmitter: Acoustic
  data exfiltration from (speakerless) air-gapped computers,'' \emph{arXiv
  preprint arXiv:1606.05915}, 2016.

\bibitem{guri2017acoustic}
------, ``Acoustic data exfiltration from speakerless air-gapped computers via
  covert hard-drive noise (diskfiltration),'' in \emph{European Symposium on
  Research in Computer Security}.\hskip 1em plus 0.5em minus 0.4em\relax
  Springer, 2017, pp. 98--115.

\bibitem{Guri2014}
M.~Guri, G.~Kedma, A.~Kachlon, and Y.~Elovici, ``{AirHopper}: Bridging the
  air-gap between isolated networks and mobile phones using radio
  frequencies,'' in \emph{Proc. 9th Int. Conf. Malicious and Unwanted Software:
  The Americas (MALWARE)}, Oct. 2014, pp. 58--67.

\bibitem{2018powerhammer}
M.~{Guri}, B.~{Zadov}, D.~{Bykhovsky}, and Y.~{Elovici}, ``{PowerHammer:
  Exfiltrating Data from Air-Gapped Computers through Power Lines},''
  \emph{ArXiv e-prints}, Apr. 2018.

\bibitem{deshotels2014inaudible}
L.~Deshotels, ``Inaudible sound as a covert channel in mobile devices.'' in
  \emph{WOOT}, 2014.

\bibitem{hanspach2014covert}
M.~Hanspach and M.~Goetz, ``On covert acoustical mesh networks in air,''
  \emph{arXiv preprint arXiv:1406.1213}, 2014.

\bibitem{carrara2014acoustic}
B.~Carrara and C.~Adams, ``On acoustic covert channels between air-gapped
  systems,'' in \emph{International Symposium on Foundations and Practice of
  Security}.\hskip 1em plus 0.5em minus 0.4em\relax Springer, 2014, pp. 3--16.

\bibitem{Krishnamurthy2018}
P.~Krishnamurthy, F.~Khorrami, R.~Karri, D.~Paul-Pena, and H.~Salehghaffari,
  ``Process-aware covert channels using physical instrumentation in
  cyber-physical systems,'' \emph{{IEEE} Transactions on Information Forensics
  and Security}, vol.~13, no.~11, pp. 2761--2771, Nov. 2018.

\bibitem{guri2016optical}
M.~Guri, O.~Hasson, G.~Kedma, and Y.~Elovici, ``{VisiSploit}: An optical
  covert-channel to leak data through an air-gap,'' in \emph{14th Annual
  Conference on Privacy, Security and Trust (PST)}.\hskip 1em plus 0.5em minus
  0.4em\relax IEEE, 2016, pp. 642--649.

\bibitem{loughry2002information}
J.~Loughry and D.~A. Umphress, ``Information leakage from optical emanations,''
  \emph{ACM Transactions on Information and System Security (TISSEC)}, vol.~5,
  no.~3, pp. 262--289, 2002.

\bibitem{lopes2017platform}
A.~C. Lopes and D.~F. Aranha, ``Platform-agnostic low-intrusion optical data
  exfiltration.'' in \emph{ICISSP}, 2017, pp. 474--480.

\bibitem{Guri2017}
M.~Guri, B.~Zadov, and Y.~Elovici, \emph{{LED-it-GO}: Leaking (a lot of) Data
  from Air-Gapped Computers via the (small) Hard Drive {LED}}.\hskip 1em plus
  0.5em minus 0.4em\relax Cham: Springer International Publishing, 2017, pp.
  161--184.

\bibitem{Loughry2018}
J.~Loughry, ``Optical {TEMPEST},'' in \emph{2018 International Symposium on
  Electromagnetic Compatibility ({EMC} {EUROPE})}.\hskip 1em plus 0.5em minus
  0.4em\relax {IEEE}, Aug. 2018.

\bibitem{Bak2019}
D.~Bak, P.~Mazurek, and D.~Oszutowska{\textendash}Mazurek, ``Optimization of
  demodulation for air{\textendash}gap data transmission based on backlight
  modulation of screen,'' in \emph{Lecture Notes in Computer Science}.\hskip
  1em plus 0.5em minus 0.4em\relax Springer International Publishing, 2019, pp.
  71--80.

\bibitem{Zhou2019}
Z.~Zhou, W.~Zhang, S.~Li, and N.~Yu, ``Potential risk of {IoT} device
  supporting {IR} remote control,'' \emph{Computer Networks}, vol. 148, pp.
  307--317, Jan. 2019.

\bibitem{Guri2019}
M.~Guri and D.~Bykhovsky, ``{aIR}-jumper: Covert air-gap
  exfiltration/infiltration via security cameras {\&} infrared ({IR}),''
  \emph{Computers {\&} Security}, vol.~82, pp. 15--29, may 2019.

\bibitem{guri2015bitwhisper}
M.~Guri, M.~Monitz, Y.~Mirski, and Y.~Elovici, ``Bitwhisper: Covert signaling
  channel between air-gapped computers using thermal manipulations,'' in
  \emph{Computer Security Foundations Symposium (CSF), 2015 IEEE 28th}.\hskip
  1em plus 0.5em minus 0.4em\relax IEEE, 2015, pp. 276--289.

\bibitem{guri2019ctrl}
M.~Guri, B.~Zadov, D.~Bykhovsky, and Y.~Elovici, ``Ctrl-alt-led: Leaking data
  from air-gapped computers via keyboard leds,'' in \emph{2019 IEEE 43rd Annual
  Computer Software and Applications Conference (COMPSAC)}, vol.~1.\hskip 1em
  plus 0.5em minus 0.4em\relax IEEE, 2019, pp. 801--810.

\bibitem{guri2018xled}
M.~Guri, B.~Zadov, A.~Daidakulov, and Y.~Elovici, ``xled: Covert data
  exfiltration from air-gapped networks via switch and router leds,'' in
  \emph{2018 16th Annual Conference on Privacy, Security and Trust
  (PST)}.\hskip 1em plus 0.5em minus 0.4em\relax IEEE, 2018, pp. 1--12.

\bibitem{Sar13}
\BIBentryALTinterwordspacing
G.~Sarah. (2013, May) How to make a computer screen invisible. dailymail.
  [Online]. Available:
  \url{http://www.dailymail.co.uk/sciencetech/article-2480089/How-make-screen-INVISIBLE-Scientist-shows-make-monitor-blank-using-3D-glasses.html}
\BIBentrySTDinterwordspacing

\bibitem{Osnos2017}
\BIBentryALTinterwordspacing
E.~Osnos, D.~Remnick, and J.~Yaffa, ``Trump, {Putin}, and the new {Cold War},''
  The New Yorker, Mar. 2017, [Online; accessed 26 August 2018]. [Online].
  Available:
  \url{https://www.newyorker.com/magazine/2017/03/06/trump-putin-and-the-new-cold-war}
\BIBentrySTDinterwordspacing

\bibitem{Tri15}
\BIBentryALTinterwordspacing
I.~Khimji. (2015, May) The malicious insider. [Online]. Available:
  \url{http://www.tripwire.com/state-of-security/security-awareness/the-malicious-insider/}
\BIBentrySTDinterwordspacing

\bibitem{Tec}
\BIBentryALTinterwordspacing
M.~Rouse. evil maid attack. [Online]. Available:
  \url{http://searchsecurity.techtarget.com/definition/evil-maid-attack}
\BIBentrySTDinterwordspacing

\bibitem{Las09}
A.~H. Lashkari, S.~Farmand, D.~O.~B. Zakaria, and D.~R. Saleh, ``Shoulder
  surfing attack in graphical password authentication,'' \emph{International
  Journal of Computer Science and Information Security}, vol.~6, no.~2, pp.
  145--154, Nov. 2009.

\bibitem{Morkel2005}
\BIBentryALTinterwordspacing
T.~Morkel, J.~H. Eloff, and M.~S. Olivier, ``An overview of image
  steganography.'' in \emph{ISSA}, 2005, pp. 1--11. [Online]. Available:
  \url{http://martinolivier.com/open/stegoverview.pdf}
\BIBentrySTDinterwordspacing

\bibitem{Yuan2012}
W.~Yuan, K.~Dana, A.~Ashok, M.~Gruteser, and N.~Mandayam, ``Dynamic and
  invisible messaging for visual {MIMO},'' in \emph{Proc. IEEE Workshop the
  Applications of Computer Vision (WACV)}, Jan. 2012, pp. 345--352.

\bibitem{Wang2014}
A.~Wang, C.~Peng, O.~Zhang, G.~Shen, and B.~Zeng, ``Inframe: Multiflexing
  full-frame visible communication channel for humans and devices,'' in
  \emph{Proceedings of the 13th ACM Workshop on Hot Topics in Networks}, ser.
  HotNets-XIII.\hskip 1em plus 0.5em minus 0.4em\relax New York, NY, USA: ACM,
  2014, pp. 23:1--23:7.

\bibitem{Li2015}
\BIBentryALTinterwordspacing
T.~Li, C.~An, X.~Xiao, A.~T. Campbell, and X.~Zhou, ``Real-time screen-camera
  communication behind any scene,'' in \emph{Proceedings of the 13th Annual
  International Conference on Mobile Systems, Applications, and Services}, ser.
  MobiSys '15.\hskip 1em plus 0.5em minus 0.4em\relax New York, NY, USA: ACM,
  2015, pp. 197--211. [Online]. Available:
  \url{http://doi.acm.org/10.1145/2742647.2742667}
\BIBentrySTDinterwordspacing

\bibitem{Kim2015}
B.~W. Kim, H.~C. Kim, and S.~Y. Jung, ``Display field communication:
  Fundamental design and performance analysis,'' \emph{Journal of Lightwave
  Technology}, vol.~33, no.~24, pp. 5269--5277, Dec. 2015.

\bibitem{Jung2018}
S.-Y. Jung, H.-C. Kim, and B.~W. Kim, ``Implementation of two-dimensional
  display field communications for enhancing the achievable data rate in
  smart-contents transmission,'' \emph{Displays}, jul 2018.

\bibitem{Wang2017}
J.~Wang, W.~Huang, and Z.~Xu, ``Demonstration of a covert camera-screen
  communication system,'' in \emph{Proc. 13th Int. Wireless Communications and
  Mobile Computing Conf. (IWCMC)}, Jun. 2017, pp. 910--915.

\bibitem{Saleh2007}
B.~E. Saleh and M.~C. Teich, \emph{Fundamentals of Photonics}, 2nd~ed.\hskip
  1em plus 0.5em minus 0.4em\relax Wiley, 2007.

\bibitem{Proakis5th}
J.~Proakis and M.~Salehi, \emph{Digital Communications}, 5th~ed.\hskip 1em plus
  0.5em minus 0.4em\relax McGraw-Hill, 2008.

\bibitem{Whe1}
\BIBentryALTinterwordspacing
J.~Liu, S.-M. Morgens, R.~C. Sumner, L.~Buschmann, Y.~Zhang, and J.~Davis,
  ``When does the hidden butterfly not flicker?'' in \emph{SIGGRAPH Asia 2014
  Technical Briefs}.\hskip 1em plus 0.5em minus 0.4em\relax ACM, 2014, p.~3.
  [Online]. Available: \url{https://graphics.soe.ucsc.edu/papers/flicker/}
\BIBentrySTDinterwordspacing

\bibitem{Ada1}
\emph{Proceedings of the National Academy of Sciences of the United Stated of
  America}.

\bibitem{Yam12}
H.~Yamamoto, S.~Farhan, S.~Motoki, and S.~S., ``Development of 480-fps led
  display by use of spatiotemporal mapping,'' 2012.

\bibitem{Placeholder1}
A.~Cheddad, J.~Condell, K.~Curran, and P.~M. Kevitt, ``Digital image
  steganography: Survey and analysis of current methods,'' vol.~90, p.
  727–752, 2010.

\bibitem{Gou91}
P.~Gouras, \emph{Color vision}.\hskip 1em plus 0.5em minus 0.4em\relax
  Elsevier, 1991, pp. 467--479.

\bibitem{Cor99}
S.~Coren, L.~M. Ward, and J.~T. Enns, ``The visual system,'' in \emph{Sensation
  and perception, 5th edition}.\hskip 1em plus 0.5em minus 0.4em\relax
  Harcourt, 1999.

\bibitem{Cor991}
------, ``Brightness and spatial frequency,'' in \emph{Sensation and
  Perception, 5th edition}.\hskip 1em plus 0.5em minus 0.4em\relax Harcourt,
  1999.

\bibitem{Cor992}
------, ``The constancies,'' in \emph{Sensation and perception, 5th
  edition}.\hskip 1em plus 0.5em minus 0.4em\relax Harcourt, 1999.

\bibitem{Cor993}
------, ``Time,'' in \emph{Sensation and perception}.\hskip 1em plus 0.5em
  minus 0.4em\relax Harcourt, 1999.

\end{thebibliography}

\end{document}